\documentclass[%
superscriptaddress,
amsmath,
amssymb,
aps,
physrev,
twocolumn,
nofootinbib,
]{revtex4-2}

\usepackage{amsmath, amsthm, amssymb}
\usepackage[]{graphicx} 
\usepackage{dcolumn} 
\usepackage[colorlinks=true, allcolors=blue]{hyperref}
\usepackage{bm} 
\usepackage[normalem]{ulem}
\usepackage{cases}
\usepackage{color}
\usepackage[dvipsnames]{xcolor}
\usepackage{algorithm}
\usepackage{algpseudocode}
\usepackage{csquotes}
\usepackage{mathtools}
\usepackage{mathrsfs}  
\usepackage{mathtools}
\usepackage{lipsum}
\usepackage{comment}
\usepackage{multirow}
\usepackage{braket}
\usepackage{pgf} 
\usepackage{siunitx} 
\usepackage{graphicx}
\usepackage{ragged2e}
\usepackage{glossaries}

\makeglossaries
\newacronym{qkd}{QKD}{quantum key distribution}
\newacronym{qber}{QBER}{quantum bit error rate}
\newacronym{dac}{DAC}{digital-to-analog converter}
\newacronym{skr}{SKR}{secret key rate}
\newacronym{snspd}{SNSPD}{superconducting nanowire single photon detector}

\newcommand{\skr}[0]{\text{SKR}}
\newcommand{\qber}[0]{\text{QBER}}

\begin{document}

\title{GHz-rate polarization-based QKD system for fiber and satellite applications}

\author{Matías~R.~Bolaños}
\affiliation{Dipartimento di Ingegneria dell'Informazione, Universit\`a degli Studi di Padova, via Gradenigo 6B, IT-35131 Padova, Italy}

\author{Edoardo~Rossi}
\affiliation{Dipartimento di Ingegneria dell'Informazione, Universit\`a degli Studi di Padova, via Gradenigo 6B, IT-35131 Padova, Italy}

\author{Federico Berra}
\affiliation{Dipartimento di Ingegneria dell'Informazione, Universit\`a degli Studi di Padova, via Gradenigo 6B, IT-35131 Padova, Italy}

\author{Alberto~De~Toni}
\affiliation{Dipartimento di Ingegneria dell'Informazione, Universit\`a degli Studi di Padova, via Gradenigo 6B, IT-35131 Padova, Italy}

\author{Ilektra~Karakosta-Amarantidou}
\affiliation{Dipartimento di Ingegneria dell'Informazione, Universit\`a degli Studi di Padova, via Gradenigo 6B, IT-35131 Padova, Italy}

\author{Daniel~C.~Lawo}
\affiliation{Dipartimento di Ingegneria dell'Informazione, Universit\`a degli Studi di Padova, via Gradenigo 6B, IT-35131 Padova, Italy}

\author{Costantino~Agnesi}
\affiliation{Dipartimento di Ingegneria dell'Informazione, Universit\`a degli Studi di Padova, via Gradenigo 6B, IT-35131 Padova, Italy}
\affiliation{Padua Quantum Technologies Research Center, Universit\`a degli Studi di Padova, via Gradenigo 6A, IT-35131 Padova, Italy}
\author{Marco~Avesani}
\affiliation{Dipartimento di Ingegneria dell'Informazione, Universit\`a degli Studi di Padova, via Gradenigo 6B, IT-35131 Padova, Italy}
\affiliation{Padua Quantum Technologies Research Center, Universit\`a degli Studi di Padova, via Gradenigo 6A, IT-35131 Padova, Italy}
\author{Andrea~Stanco}
\affiliation{Dipartimento di Ingegneria dell'Informazione, Universit\`a degli Studi di Padova, via Gradenigo 6B, IT-35131 Padova, Italy}
\affiliation{Padua Quantum Technologies Research Center, Universit\`a degli Studi di Padova, via Gradenigo 6A, IT-35131 Padova, Italy}
\author{Francesco~Vedovato}
\affiliation{Dipartimento di Ingegneria dell'Informazione, Universit\`a degli Studi di Padova, via Gradenigo 6B, IT-35131 Padova, Italy}
\affiliation{Padua Quantum Technologies Research Center, Universit\`a degli Studi di Padova, via Gradenigo 6A, IT-35131 Padova, Italy}
\author{Paolo~Villoresi}
\affiliation{Dipartimento di Ingegneria dell'Informazione, Universit\`a degli Studi di Padova, via Gradenigo 6B, IT-35131 Padova, Italy}
\affiliation{Padua Quantum Technologies Research Center, Universit\`a degli Studi di Padova, via Gradenigo 6A, IT-35131 Padova, Italy}
\author{Giuseppe~Vallone}
\affiliation{Dipartimento di Ingegneria dell'Informazione, Universit\`a degli Studi di Padova, via Gradenigo 6B, IT-35131 Padova, Italy}
\affiliation{Padua Quantum Technologies Research Center, Universit\`a degli Studi di Padova, via Gradenigo 6A, IT-35131 Padova, Italy}
\date{\today}

\begin{abstract}

\Gls{qkd} leverages the principles of quantum mechanics to exchange a secret key between two parties. 
Despite its promising features, \gls{qkd} also faces several practical challenges such as transmission loss, noise in quantum channels and finite key size effects. Addressing these issues is crucial for the large-scale deployment of \gls{qkd} in fiber and satellite networks.

In this paper, we present a 1550 nm \gls{qkd} system realizing the efficient-BB84 protocol and based on the iPOGNAC scheme.
The system achieved repetition rates up to 1.5~GHz and showed an intrinsic QBER of $\sim 0.4\%$. 
The system was first tested on a laboratory fiber link and then on an intermodal link in the field, consisting of both deployed fiber and a 620 m free-space channel. 
The experiment was performed in daylight conditions, exploiting the Qubit4Sync synchronization protocol. 
With this trial, we achieved a new benchmark for free-space BB84 \gls{qkd} systems by generating a sustained \gls{skr} above 1~Mb/s for 1 hour.
Finally, exploiting a recently discovered finite-size bound, we achieved a secure key rate of about 10 Mb/s at low losses (5 dB), and around 6.5~kb/s in the high-loss (38.5 dB), low block length ($N=10^4$) regime.
The latter results demonstrate the system's suitability for highly lossy and time-constrained scenarios such as \gls{qkd} from low Earth orbit satellites.

\end{abstract}

\maketitle
\glsresetall
\section{Introduction}

In the domain of secure communication, \gls{qkd} stands out as a crucial tool for ensuring the safe exchange of sensitive information between distant parties \cite{bennettQuantumCryptographyPublic2014, gisinQuantumCryptography2002}. 
Unlike traditional key exchange methods, \gls{qkd} relies on the laws of quantum mechanics to create secret keys that are designed to avoid eavesdropping. 
When coupled with an information-theoretic secure symmetric encryption scheme (e.g. one-time pad), \gls{qkd} offers unconditional security, making it a valuable asset for safeguarding sensitive data. 
Protocols such as BB84~\cite{bennettQuantumCryptographyPublic2014} and its decoy-state variants~\cite{PhysRevLett.94.230504} have been demonstrated over increasingly long distances and integrated into field-deployable systems, particularly for fiber-based implementations~ \cite{lucamariniOvercomingRateDistance2018, Wang2022}.

While optical fibers suffer from exponential transmission losses that fundamentally limit \gls{qkd} to a few hundred kilometers without quantum repeaters \cite{Pirandola2017, Martin2021,  deForgesdeParny2023}, free-space channels offer a much more favorable distance scaling, with losses following a square-law relationship \cite{deForgesdeParny2023, Yin2017, Ren2017}.
Recent terrestrial free-space \gls{qkd} trials have demonstrated feasibility in both discrete-variable (DV) and continuous-variable (CV) regimes, but their achievable secure key rates remain limited under real-world conditions. For instance, Cocchi et al.~\cite{Cocchi:25} reported a time-bin DV-QKD system operating at 595 MHz over a 500~m urban link, where an optical clock signal was multiplexed with the quantum one for synchronization purposes, achieving a finite-key \gls{skr} of approximately 40~kb/s under daylight. In the CV regime, Zheng et al.~\cite{zheng2025free} demonstrated free-space \gls{qkd} over 860~m with an asymptotic \gls{skr} around 300~kb/s, which reduced to around 40~kb/s after taking into account finite-size effects.
However, ground-to-ground free-space links are severely constrained by line-of-sight availability and atmospheric inhomogeneities, which become more detrimental closer to sea level and as the distance between the communicating parties increases above one kilometer~\cite{Pirandola2021satellite, Avesani-fieldtrial}. Taking this into account, the vision of a truly global quantum-secure network requires solutions that go beyond terrestrial links.
Satellite-based \gls{qkd} overcomes these limitations by placing the transmitter or receiver above the atmosphere, enabling line-of-sight links to widely separated ground stations and making intercontinental quantum communication possible.
In particular, the scenario where the transmitter is placed above the atmosphere is preferred, as the receiver contains one or more single photon detectors, which not only are not expected to function normally in above-atmosphere conditions, but are also bulk components not targeted for size-constrained scenarios like a satellite.
On the other hand, the transmitter can consist mostly of a laser source and electro-optic components, both of which have space-qualified versions commercially available.

Satellite-to-ground links require extra effort with respect to fiber or ground-to-ground free-space links, particularly because the key generation time is limited by the satellite passage over the ground station. 
Coupling that with high link losses between the parties due to longer channels, satellite \gls{qkd} sources should work at ultrafast repetition rates to maximize key generation. 
Moreover, atmospheric turbulence creates rapid changes on the link's efficiency, which when using large block sizes for a protocol run, can have a detrimental impact on the \gls{skr}.

In this work, we present a high-speed (1.5~GHz) and low-error three-state polarization-encoded 1550 nm \gls{qkd} system that is suitable for both fiber and satellite communication. 
The source includes two \textit{iPOGNAC}s~\cite{Avesani:iPOGNAC} in the {\it symmetric} configuration~\cite{Berra2025General} used for intensity modulation and polarization encoding respectively, as proposed in \cite{Berra2023_2ipognac}. 
The system was first tested under laboratory conditions with a low-loss fiber optical link, where it achieved a stable secure key rate over an hour-long measurement, followed by two field-trial tests on an intermodal link consisting of two fiber paths and a 620~m free-space channel. 
The intermodal link was first operated at its nominal conditions, achieving what is, to our knowledge, a new benchmark for secure key rates in metropolitan free-space \gls{qkd} under daylight, taking both DV and CV regimes into account. 
Then, extra losses were added in the channel to emulate similar conditions to a satellite-ground link, reaching a total of 38.5~dB, where positive secure key was extracted every $\sim0.3$~s by taking a finite block size of $N=10^4$.
For the intermodal link experiments, the synchronization between the \gls{qkd} transmitter and receiver was performed via Qubit4Sync \cite{Calderaro2020}, representing the implementation at the highest repetition rate of a qubit-based syncronization method.

\begin{figure*}
    \centering
    \includegraphics[width=0.97\textwidth]{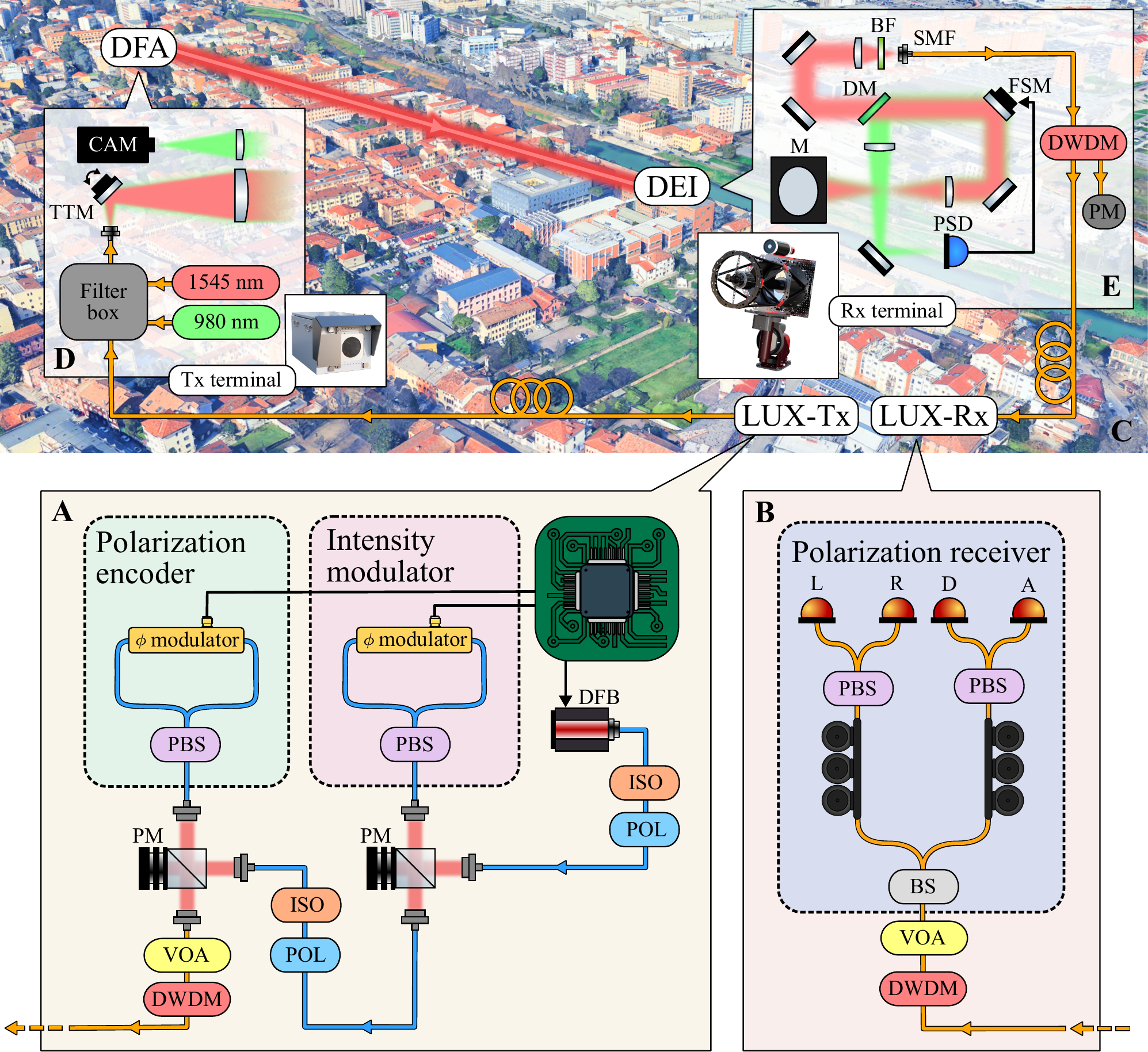}
    
    \caption{\justifying{\textbf{(A)} The \gls{qkd} source is composed by a gain-switched DFB laser generating optical pulses that go through an isolator (ISO), and a polarizer (POL) that projects the light into the $\ket{D}$ polarization. 
    The polarized pulses are injected into a free-space-coupled beam-splitter (BS): the beam gets reflected into an iPOGNAC modulator, consisting of a polarizing beamsplitter (PBS) and a phase-modulator ($\phi$-Mod), and travels back to be transmitted.
    The output states are once again injected to an isolator and a $\ket{D}$-aligned polarizer, effectively attenuating the optical pulses by a factor of $\cos^2(\Delta\phi)$, which  are then used as input states to a second iPOGNAC modulator used for polarization encoding. 
    The transmission port of the free-space-coupled BS for this modulation stage is coupled to a single-mode fiber (SMF), including a variable optical attenuator (VOA) to control the mean photon number of the source, and a dense wavelength-division multiplexer (DWDM) to spectrally filter the quantum signal.
    \textbf{(B)} The receiver consists of a DWDM to remove off-band noise introduced in the free-space channel, followed by a VOA to control channel losses, a fiber BS, two polarization-controllers (PC), two PBSs and four detectors, one for each polarization state (namely L, R, D, A). \textbf{(C)} Aerial view of the 620 m free-space channel between DFA and DEI in Padova. The \gls{qkd} source and receiver are located in the same laboratory, but we differentiate the Tx components from the Rx ones as LUX-Tx and LUX-Rx. Data from Google Earth [\textcopyright2025 Google]. \textbf{(D)} Schematic of the Tx terminal at DFA, connected to LUX-Tx through a deployed fiber network. It is also shown the render of the 3D model of the enclosure hosting the optical terminal. \textbf{(E)} Schematic of the optical components at Rx terminal. It is also shown the render of the 3D model of the receiving telescope. TTM: tip-tilt mirror; CAM: camera; M: 2-inch mirror; FSM: fast-steering mirror; DM: dichroic mirror; BF: bulk filter; SMF: single-mode fiber; PSD: position-sensitive-detector; PM: power meter.
    }}
    \label{fig:quangoscheme}
\end{figure*}

\section{Methods}

In this section, we present the essential technology used to conduct the experiments of GHz-rate polarization-based \gls{qkd}. First, we elaborate on the design of the $1.5$~GHz source, which implements a three-state, one-decoy BB84 \gls{qkd} protocol encoded in polarization and working at a wavelength of $1550$~nm. Secondly, we describe the free-space terminals that we interfaced with the source to perform the intermodal test over a $620$~m free-space link.

\subsection{Source design and implementation}

We engineered a \gls{qkd} source and receiver of polarization-encoded states at 1550 nm with the capability to execute the efficient 3-states 1-decoy BB84 protocol, described in \cite{Grunenfelder_2018}. 
As illustrated in Fig.~\ref{fig:quangoscheme}~(A), the source was realized with a gain-switched distributed-feedback (DFB) 1550 nm laser followed by a sequence of two iPOGNAC encoders \cite{Avesani:iPOGNAC, Berra2025General}, consisting of a Sagnac loop modulated by an electro-optical phase modulator (EOM). 
 
As shown in \cite{Avesani:iPOGNAC}, when the input polarization state for the iPOGNAC encoder is $\ket{D}$, this system is capable of creating any polarization state on the equator of the Bloch Sphere such that $\ket{\Psi}=\ket{H}+e^{-i\Delta\phi}\ket{V}$. 
Thus, by combining one of the iPOGNAC encoders with a $\ket{D}$-aligned polarizer at its output, it is possible to attenuate incoming light pulses by a factor of $\cos^2(\Delta\phi)$, allowing the system to encode decoy states \cite{Berra2023_2ipognac}. 
The attenuated pulses then go through a second iPOGNAC stage, in which the polarization is encoded into one of the three states required for the protocol: $\ket{D}$, $\ket{L}$ and $\ket{R}$, corresponding to diagonal, left-handed circular, and right-handed circular polarization states respectively.
The encoded signal was then sent through an optical channel (i.e. fiber or free-space) to a standard polarization receiver (Fig.~\ref{fig:quangoscheme}~(B)) consisting of a balanced beamsplitter, which performs the random basis choice, with each of its outputs followed by a polarization controller and a polarizing beam-splitter.
The polarization controllers were set such that one arm of the beam-splitter projected the incoming photons on the key basis $Y = \{\ket{L}, \ket{R}\}$, while the other on the check basis $X = \{\ket{D}, \ket{A}\}$.
 
To achieve a repetition rate in the gigahertz range, both iPOGNAC encoders were built in the symmetric configuration, following the methods presented in \cite{Berra2025General}, particularly using the \textit{balanced modulation} scheme.
Although this modulation scheme limits the maximum repetition rate, the zero-average voltage property of the electrical signals driving all symbols makes the state encoding more resilient when using AC-coupled RF amplifiers, as the DC component of any random sequence is constant and equal to zero.
To control all the electrical signals used for the EOMs on both the iPOGNAC encoders and the pulsed laser, an \textit{Ultrascale+} based system-on-a-chip (SoC) equipped with a \gls{dac} running at $6$~GSa/s was used.

\subsection{Description of the free-space terminals}
\label{sec:fsDescription}

For the free-space testbed, we located our optical terminals in two buildings of the University, both in the city of Padova. As shown in Fig.~\ref{fig:quangoscheme}~(C), the transmitter terminal (Tx) was installed at the Department of Physics and Astronomy (DFA), while the receiver terminal (Rx) at the Department of Information Engineering (DEI). 
The \gls{qkd} source and receiver were placed in the same room at the Luxor laboratories (LUX), but for readability purposes we differentiate the location of the source from the receiver one as LUX-Tx and LUX-Rx. 
After the propagation in fiber from LUX-Tx to DFA, characterized by 3.75~dB of losses, the signal further propagates in the 620~m free-space channel up to DEI, connected through a deployed fiber of losses 4.24~dB to the \gls{qkd} receivers at LUX-Rx.
To maintain high channel efficiency, we integrated two lasers into the pointing, acquisition and tracking (PAT) architecture of the Tx terminal. A 980~nm laser serves as a first-stage alignment by feeding the fast-steering mirror (FSM) at the Rx terminal, while a 1545.32~nm laser (C40 of the ITU grid) is exploited to monitor channel efficiency during the field trial. This dual-wavelength approach allows the system to simultaneously maintain alignment and provide a continuous metric of link performance via SMF power coupling.
These wavelengths are then combined with the quantum signal by means of a set of dense wavelength-division multiplexers (DWDMs) enclosed in a filter box.
Once the signals are multiplexed, they propagate through an SMF connected to a linear stage positioned at the rear focal plane of a 2-inch lens, which is part of the Tx optical terminal. 
Using the C40 monitoring laser as a reference, a second-stage alignment is subsequently performed through a remote-controlled tip-tilt mirror that is placed between the focal plane and the lens, allowing for precision of approximately $2~\mu\mathrm{rad}$~\cite{GuerriniBachelor2022}.
After the lens, the signals are sent through free-space as collimated beams of waist $W_0 = 25$ mm, through the enclosure's optical window of diameter $150$ mm, minimizing the reflection losses for the exploited wavelengths.
The overall losses from the Tx input fiber to the output of the optical window are 1.2~dB. 
The Tx terminal is completely controllable remotely, and allows alignment with the Rx by using an alt-azimuth mount, and a camera (CAM) with a full-angle field of view (FOV) of $8 \times 6.4$~mrad$^2$.

At the receiver side, the Rx is housed within the optical ground station (OGS) located on the roof of DEI (Fig.~\ref{fig:quangoscheme}~(E)). 
The terminal, an $f/8$ Ritchey-Chrétien telescope, is equipped with a $600 \times 600$ mm$^2$ breadboard mounted at its rear. 
This breadboard hosts the optical path for SMF injection, reachable through a 2-inch mirror (M) tilted by 45 degrees with respect to the telescope's optical axis.
On the breadboard, a lens with $100$ mm focal length collimates the received beam after the primary focus, forming a pupil of dimension $12.5$ mm on the surface of a 1-inch fast-steering mirror (FSM), which is included to address any tip-tilt fluctuation caused by atmospheric turbulence. 
Following the reflection off the FSM, a dichroic mirror (DM) is placed along the path in order to separate the C-band signals from the $980$~nm beam. 
The $980$~nm beam is detected by a silicon position sensitive detector (PSD) after going through a 300 mm focal length lens, which will record the displacements of the centroid on the focal plane. 
This information is used to adjust the voltage applied on the FSM's actuators, minimizing centroid fluctuations.
The C-band signals are instead coupled into an FC/UPC-terminated SMF, located after a bulk bandpass filter (BF). 
The filter is centered at $1550$ nm, with a bandwidth of 40 nm full-width-at-half-maximum (FWHM), and is placed before the lens used to couple the signal into the SMF. 
The focal length of this lens is chosen to guarantee that the effective focal length of the system satisfies $f_{\rm eff} = \pi\varnothing {\rm MFD}/\left(4\lambda \beta\right)$ \cite{Scriminich_2022}, where MFD is the mode-field diameter of the SMF ($10.4$ $\mu$m), $\beta \approx 1.12$, and $\varnothing$ is the value of the diameter of the beam in front of the SMF.
Once in the fiber, the C-band signals are separated through a set of $100$ GHz filters. 
The C40 signal is connected to a fiber coupled power-meter (PM), with which the injected power is logged to estimate the channel losses $\eta_{\rm CH}$, while the \gls{qkd} signal is sent to Bob at LUX-Rx for the quantum measurements.

\section{Results and Discussion}
In this section we evaluate the system performance under two configurations: a controlled laboratory fiber link and an intermodal link consisting of two deployed fiber links connected through a 620~m urban free-space optical link. 
In both cases, the same GHz-rate, polarization-encoded \gls{qkd} source implementing the three-state, one-decoy BB84 protocol was employed.
The performance was studied firstly by characterizing the system stability and maximum achievable \gls{skr} over the laboratory fiber link. 
Then, we show how the free-space optical channel was optimized to minimize losses by using a fast steering mirror, managing to obtain a new benchmark in secure key generation for \gls{qkd} protocols implemented over free-space links.

\subsection{Laboratory fiber link}\label{sec:fiberlink}

\begin{figure}
    \centering
    \includegraphics{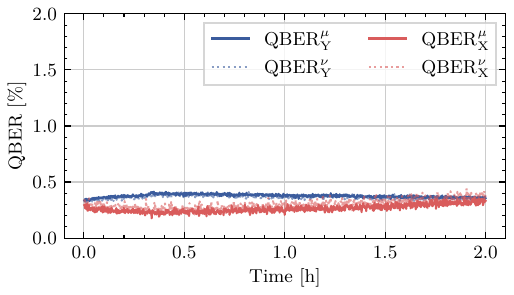}
    \caption{\justifying{Measured $\qber_B^k$ over time for a 1024-symbol pseudo-random sequence, where $B \in \{Y,X\}$ denotes the key and check bases, respectively, and $k \in \{\mu,\nu\}$ the signal and decoy states.}}
    \label{fig:qber-stability-2hs}
\end{figure}

To validate the performance of the source for \gls{qkd} applications, a fixed pseudo-random sequence of 1024 symbols was generated to emulate the random generation of quantum states for a three-state, one-decoy BB84 implementation. 
As an initial test, the source was operated continuously at a repetition rate of 1~GHz for two hours, without any active or passive stabilization applied to its components. 
To prevent saturation effects on the \glspl{snspd}, a total attenuation of 26~dB was introduced in the channel. 
Throughout the run, the \gls{qber} was recorded for both Y and X bases, at the signal ($\mu$) and decoy ($\nu$) intensity levels. 
Over the full duration, a $\qber_Y = (0.38 \pm 0.01)\%$ for the key basis and $\qber_X = (0.27 \pm 0.03)\%$ for the check basis were obtained (Fig.~\ref{fig:qber-stability-2hs}). 
We note that the slow temporal drift observed in the \gls{qber} is consistent with polarization fluctuations induced by SMFs in the receiver path.

\begin{figure}
    \centering
    \includegraphics{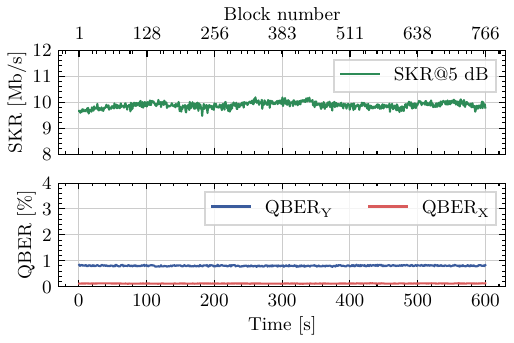}
    \caption{\justifying{Estimated \gls{skr} at $5$~dB channel losses for the three-state one-decoy BB84 protocol at $1.5$~GHz repetition rate (green line, upper axis). In the lower axis, the measured \gls{qber} for both key and check bases (in red and blue respectively).}}
    \label{fig:skr-1.5GHz}
\end{figure}

To explore the system's performance limit, the source was then operated at a repetition rate of $1.5$~GHz, close to the theoretical maximum of $1.53$~GHz achievable with balanced modulation and a $6$~GSa/s \gls{dac}~\cite{Berra2025General}. 
For this test, the protocol parameters were controlled to be $\mu = 0.28$, $\nu/\mu \approx 0.44$, $p_Z = 0.9$, and $p_\mu = 0.5$. 
The \gls{qber} and \gls{skr} were estimated over a 10-minute acquisition with a channel loss of 5~dB to avoid detector saturation. 
To estimate the \gls{skr}, the Chernoff statistical bound was used~\cite{mannalath2025sharp}, obtaining an average $\skr = 9.89 \pm 0.11$~Mb/s, reaching a maximum of $\skr = 10.19$~Mb/s when using a realistic block length of $N=10^7$ (Fig.~\ref{fig:skr-1.5GHz}). 

\begin{figure}
    \centering
    \includegraphics{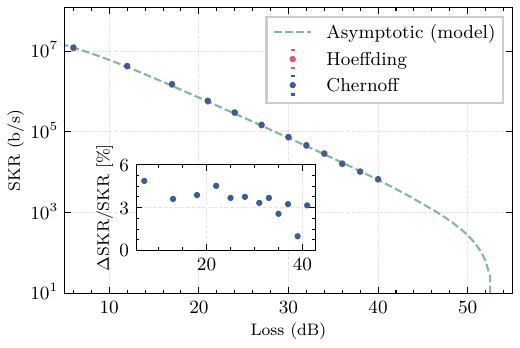}
    \caption{\justifying{Estimated \gls{skr} as a function of channel attenuation. 
    The dashed line indicates the asymptotic limit, while the red and blue points correspond to finite-size analyses using the Hoeffding and Chernoff bounds, respectively. 
    Inset: Difference in the estimated key rate between the Chernoff and Hoeffding bounds.
    }
}
    \label{fig:skr-vs-losses}
\end{figure}

Then, by introducing a variable optical attenuator between the source and receiver, the finite-size \gls{skr} was estimated for a range of channel attenuation values. 
For each loss value, at least 5 full sifted key blocks were acquired for statistical significance. 
The \gls{skr} was calculated using both Hoeffding~\cite{Rusca2018} and Chernoff~\cite{mannalath2025sharp} statistical bounds. 
As expected, the Chernoff bounds consistently yielded higher \gls{skr} values than the Hoeffding bounds (see inset of Fig.~\ref{fig:skr-vs-losses}), in agreement with the theoretical predictions of Ref.~\cite{mannalath2025sharp}. 
We note that the estimated \gls{skr} follows the expected behaviour as a function of the channel losses, showing that the source is capable of generating positive secure key rate even at 52~dB of channel losses. 
This result, combined with the newly found statistical bound capable of generating secret key at smaller block sizes, confirms the suitability of this system for satellite-based \gls{qkd} links.

\subsection{Intermodal link}
\label{sec:fsResults}
The field trial has been conducted on November 7th, 2025, over the intermodal optical path combining fiber and free-space segments described in section~\ref{sec:fsDescription}. 

For the free-space link, the channel efficiency $\eta_{\rm CH}$ was defined as the end-to-end loss from the exit of the transmitter enclosure’s optical window to the optical power detected by the fiber-coupled power meter after the DWDM demultiplexer at the receiver (Fig.~\ref{fig:quangoscheme}(E)).
To provide a physically meaningful description of the horizontal channel, we can factorize the total transmission efficiency into three independent contributions:
\begin{equation}
    \eta_{\rm CH} = \eta_{\rm F} \eta_{\rm Optics}^{\rm Rx} \eta_{\rm SMF} \ , 
    \label{eq:etaCh}
\end{equation}
where $\eta_{\rm F}$ accounts for all the losses up to the receiver's focal plane, like atmospheric absorption and telescope collection efficiency; $\eta_{\rm Optics}^{\rm Rx}$ describes insertion losses introduced by the receiving optical elements; and $\eta_{\rm SMF}$ captures the coupling efficiency into the SMF.  
The first term $\eta_{\rm F}$, was obtained by comparing the optical power at the transmitter with the power measured at the primary focal plane of the receiving telescope, yielding an average value equivalent to $2$~dB  of losses under the conditions of the trial.
The receiver-side optical transmission efficiency $\eta_{\rm Optics}^{\rm Rx}$, arise from reflections and non-ideal coatings of some optical components, and from the telescope's focal plane to the plane in front of the SMF we measured $1.9$ dB of losses.
The coupling efficiency $\eta_{\rm SMF}$ was measured by comparing the optical power in front of the fiber and the one measured by the fiber-based PM into the SMF.

\begin{figure}
    \centering
    \includegraphics{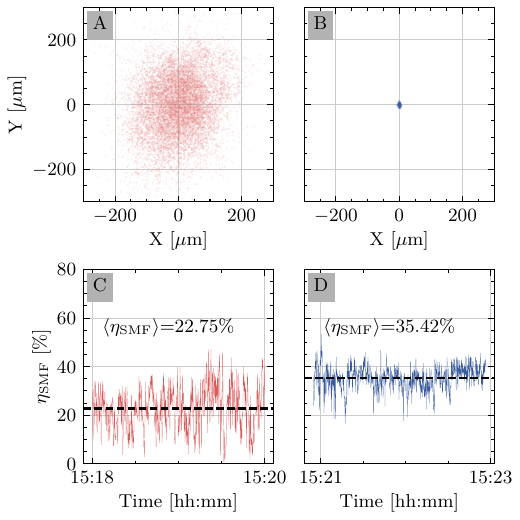}
    \caption{\justifying{Centroids displacement on the PSD when \textbf{(A)} FSM-OFF and \textbf{(B)} FSM-ON.
    Resulting system coupling efficiency with \textbf{(C)} no correction and \textbf{(D)} correcting the effect of turbulence.}}
    \label{fig:FreeSpaceResults}
\end{figure}

Free-space links are always affected by atmospheric turbulence, which manifests as beam displacement in the focal plane and as additional aberrations on the received wavefront. 
In our testbed, Fig.~\ref{fig:FreeSpaceResults}~(A) shows how turbulence induced a noticeable displacement of the coarse beam on the PSD focal plane. 
This displacement corresponds to an angle-of-arrival (AoA) fluctuation at the telescope aperture of approximately $9~\mu\mathrm{rad}$, resulting in an SMF coupling efficiency of $22.75\ \%$ (equivalent to $6.43$ dB of losses), as reported in Fig.~\ref{fig:FreeSpaceResults}~(C). 
By closing the loop on the FSM, the beam fluctuations in the focal plane are significantly reduced, as shown in Fig.~\ref{fig:FreeSpaceResults}~(B). 
In this case, the residual AoA is lowered to $0.2~\mu\mathrm{rad}$ at the telescope aperture, with a corresponding increase in SMF coupling to $\eta_{\rm SMF} = 35.42$\% (corresponding to 4.5~dB of losses), as illustrated in Fig.~\ref{fig:FreeSpaceResults}~(D).
This stabilization enhances the probability that the beam falls within the optimal coupling region, leading to a measurable improvement of roughly $2$ dB in fiber coupling and resulting in overall channel losses of 8.4~dB.

\begin{figure}
    \centering
    \includegraphics{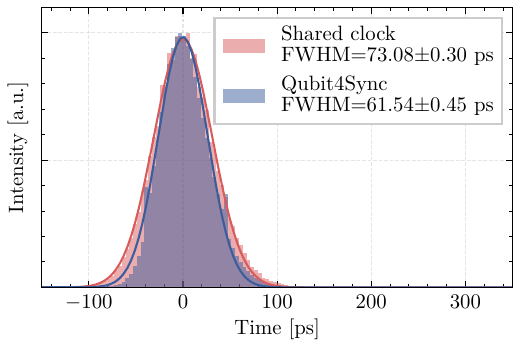}
    \caption{\justifying{Detection jitter for optical pulse when sharing a clock between Alice and Bob (in red) and when using Qubit4Sync for synchronizing both parties (in blue). 
    In solid lines, the corresponding Gaussian fit for both datasets.
    We note that the detection jitter was reduced when utilizing Qubit4Sync.}}
    \label{fig:jitter-comparison}
\end{figure}

For free-space tests, we implemented the Qubit4Sync synchronization protocol \cite{Calderaro2020} that exploits the detections of the \gls{qkd} signal to synchronize the trasnmitting and receiving device.
To measure the quality of  synchronization, we compared the optical pulse width obtained when using Qubit4Sync with the one obtained when sharing a 10 MHz clock through a direct coaxial cable. 
We note that up to now the Qubit4Sync was only tested at low repetition rate \cite{AgnesiSimpleQKD, AvesaniField} and its capabilities for high speed \gls{qkd} systems have never been demonstrated.
As shown in Fig.~\ref{fig:jitter-comparison}, Qubit4Sync not only allowed to synchronize the devices but also provides superior timing precision, outperforming the time-tagger’s internal PLL and reducing the effective detection jitter.

\begin{figure}
    \centering
    \includegraphics{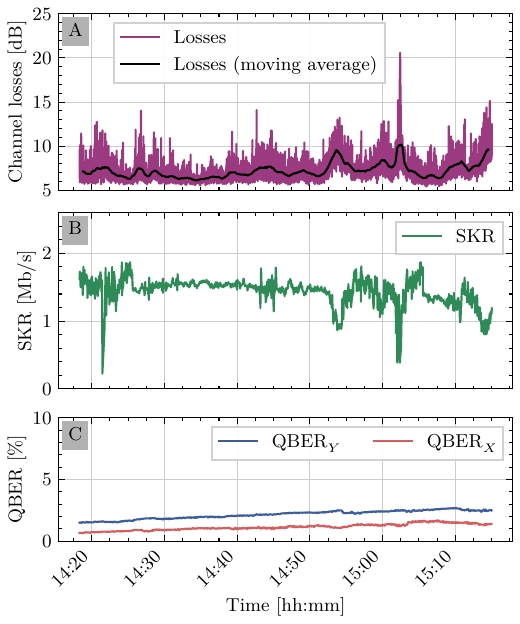}
    \caption{\justifying{\textbf{(A)} Channel losses measured between the output of the window at the Tx terminal and the power injected in the SMF measured after the DWDM at Rx terminal. \textbf{(B)} Finite-size \gls{skr} estimated using a block length of $N=1\times10^7$, and \textbf{(C)} \gls{qber} measured for both the key and check bases for the duration of the demonstration (in red and blue respectively).}}
    \label{fig:freespace-qkd}
\end{figure}

Two distinct experiments were conducted over a 620 m free-space optical link between the DFA and DEI buildings to assess the performance of the GHz polarization-based \gls{qkd} source.
In the first experiment (Fig.~\ref{fig:freespace-qkd}), the system was operated under nominal link conditions, corresponding to total average channel losses of $17.5$~dB from source output (LUX-Tx) to detector input (LUX-Rx). A continuous one-hour \gls{qkd} session was performed under daylight, using parameters adapted for this loss regime: $p_\mu = 0.7$, $p_Z = 0.9$, $\mu = 0.5$, and $\nu = 0.13$. 
At the start of the session, the measured error rates were $\qber_Y = 1.55 \pm 0.03\%$ and $\qber_X = 0.73 \pm 0.03\%$\footnote{The slightly elevated $\qber_X$ compared to Section~\ref{sec:fiberlink} is attributed to a minor electrical misalignment at the source.}. 
These values exhibited a slow temporal increase ($d\qber/dt = 2.7\times 10^{-4}\%\mathrm{/s}$), primarily due to polarization drift in the fiber segment.
Using a block size of $N = 10^7$, the system achieved a sustained \gls{skr} exceeding $1$~Mb/s, with an average of $\skr = 1.44 \pm 0.18$~Mb/s, establishing what is, to our knowledge, the highest demonstrated free-space secure key rate taking finite-size effects into account over this distance in daylight conditions.

\begin{figure}
    \centering
    \includegraphics{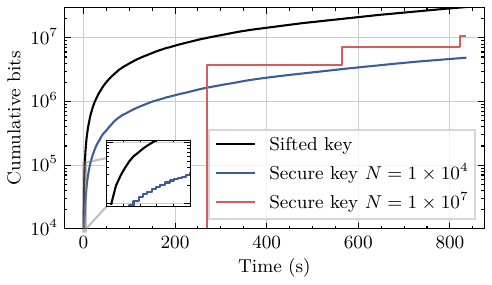}
    \caption{Cumulative bits obtained for the 15-minute measurement: Raw sifted bits (in black), secure key with block size $N=1\times10^4$ (in blue) and secure key with block size $N=1\times10^7$ (in red). Inset: Zoomed-in plot on the first 5 seconds of protocol runtime to highlight the discretization on the secure key packets over time.}
    \label{fig:cumulative-bits}
\end{figure}

In a second experiment, we emulated a satellite-to-ground attenuation regime by inserting an additional 21~dB of loss at the receiver, resulting in an effective channel loss of $\sim 38.5$~dB. 
Over a 15-minute continuous run, the system maintained stable error rates of $\qber_Y = 1.46 \pm 0.04$\% and $\qber_X = 3.24 \pm 0.20$\%. With $N = 10^7$, the corresponding \gls{skr} was $\skr = 11.69 \pm 2.76$~kb/s. 
To evaluate finite-size effects on the security of the key, we applied the finite-key bounds of Mannalath \emph{et al.}~\cite{mannalath2025sharp}. 
Additionally, to evaluate performance under the reduced acquisition times typical of low-Earth-orbit passes, a block size of $N = 10^4$ was used.
Even under such small block lengths, the system produced positive secure keys on every round of the protocol, obtaining a positive secure key rate of $\skr = 6.54 \pm 1.60$~kb/s. 
It is worth noting that, even though the \gls{skr} value is lower when comparing with using $N=10^7$, the time required to accumulate a block of key $t_{\rm key}$ is severely reduced, achieving secure keys every $t_{\rm key}=0.27\pm0.05$~s (Fig.~\ref{fig:cumulative-bits}). 

These results highlight the robustness and adaptability of the platform, confirming its suitability for both moderate-loss terrestrial links and high-loss space-communication scenarios.

\section{Conclusions}

In this work, we have successfully developed and tested a complete, high-performance Discrete-Variable \gls{qkd} system operating at repetition rates up to $1.5$~GHz. 
The system, based on polarization-encoded states at 1550 nm, exploiting the \textit{iPOGNAC} in its symmetric configuration and implementing the efficient 3-state 1-decoy BB84 protocol, was tested over both a stable fiber link and an intermodal optical channel, consisting of a 620 m urban free-space link and 2 km of fiber link. 
For all intermodal link tests, the Qubit4Sync synchronization method was used, for the first time in such high-rate \gls{qkd} setting,  achieving precise synchronization using the very same qubit signal without requiring an extra communication channel, thus simplifying the setup.

In laboratory fiber tests, the system demonstrated exceptional intrinsic stability, maintaining a low average \gls{qber} of $\sim 0.4\%$ over two hours of continuous operation. 
When pushing the source to its performance limits at $R=1.5$~GHz, an average $\skr=9.89\pm0.11$~Mb/s was estimated for $5$~dB channel losses using a realistic block length of $N=10^7$.

The source also behaved well over the intermodal link, where through the implementation of an FSM, we actively compensated the atmospheric turbulence present along the free-space channel, reducing the AoA fluctuations of the beam.
This enabled the system to achieve a sustained \gls{skr} exceeding 1 Mb/s with a stable key-basis \gls{qber} of $\sim 1.5\%$, even during daylight hours, establishing a new performance benchmark for free-space \gls{qkd}, compared to both DV and CV \gls{qkd} implemented in free-space links. 
Furthermore, we demonstrated the system's robustness by emulating a high-loss, satellite-to-ground link scenario. By introducing calibrated attenuation to reach a total channel loss of $\sim 38.5$~dB, the system successfully generated a secure key at a rate around 11.7~kb/s. 

The application of advanced finite-size security analyses confirmed that a positive key rate could be extracted even with smaller data blocks, obtaining 6.5~kb/s at a block length of $N=10^4$ at 38.5~dB channel losses, thus highlighting the system's suitability for time-constrained scenarios such as low-Earth-orbit satellite passes. 
It is worth noting that under these tight constraints, a block of secure key was distilled approximately every 0.3~s.

The optical setup was designed with fiber-based components that can be replaced with their space qualified counterpart for satellite operations. 
In this work, the electrical signals were driven using a Zynq Ultrascale+ development board with no space qualified counterpart, which could be replaced by a space qualified Zynq Ultrascale+ (Kintex XQR) coupled with a space qualified external digital-to-analog converter.

These results collectively validate our platform as a robust, high-performance, and operationally flexible solution, paving the way for future metropolitan and satellite-based quantum communication networks.

\section*{Author contributions}

F.B., M.R.B., A.D.T., C.A., M.A., G.V. and P.V. designed and implemented the optical setup, including source and receiver. 
M.R.B., D.L., A.S. designed and implemented the electrical driving system on the FPGA development board. 
E.R., I.K.A., F.V. and P.V. designed, implemented, and managed the free-space terminals.
M.R.B. and E.R. performed the experiments both in laboratory and field-trial.
All authors contributed on the writing and review of the manuscript.

\section*{Acknowledgements}
\label{sec:ack}
M.R.B. acknowledges support from the European Union’s Horizon Europe Framework Programme under the Marie Sklodowska Curie Grant No. 101072637, Project Quantum Safe Internet (QSI).
This work was (partially) supported by the ICSC—Centro Nazionale di Ricerca in High Performance Computing, Big Data and Quantum Computing, funded by European Union—NextGenerationEU

The authors would like to acknowledge A. Guerrini for the support on the design of the optical transmitter, I. Rossatelli of DFA for the technical support, and F.Bettini and F.Luise of DEI for the collaboration and support at the department.

\bibliography{references}

\end{document}